\documentclass{article}
\usepackage{setspace}

\usepackage{spconf,amsmath,graphicx}
\usepackage{multirow}
\usepackage{enumitem, kantlipsum}

\title{BINAURAL SOUND SOURCE LOCALIZATION USING A HYBRID TIME AND FREQUENCY DOMAIN MODEL}

\name{Gil Geva$^{\star}$, Olivier Warusfel$^{\dagger}$, Shlomo Dubnov$^{\ddagger}$, Tammuz Dubnov$^{\star}$, Amir Amedi$^{\star}$,  Yacov Hel-Or$^{\star}$}

\address{$^{\star}$ Reichman University, Herzliya, Israel \ $^{\ddagger}$University of California, San Diego \\ $^{\dagger}$STMS, IRCAM - CNRS - Sorbonne Université - Ministère de la Culture, Paris, France}

\begin{document}
%
\maketitle

\begin{abstract}

This paper introduces a new approach to sound source localization using head-related transfer function (HRTF) characteristics, which enable precise full-sphere localization from raw data. While previous research focused primarily on using extensive microphone arrays in the frontal plane, this arrangement often encountered limitations in accuracy and robustness when dealing with smaller microphone arrays.
Our model proposes using both time and frequency domain for sound source localization while utilizing Deep Learning (DL) approach. The performance of our proposed model, surpasses the current state-of-the-art results. Specifically, it boasts an average angular error of $0.24^\circ$ and an average Euclidean distance of $0.01$ meters, while the known state-of-the-art gives average angular error of $19.07^\circ$ and average Euclidean distance of $1.08$ meters.
This level of accuracy is of paramount importance for a wide range of applications, including robotics, virtual reality, and aiding individuals with cochlear implants (CI).

\end{abstract}
\vspace{-0.3em}

\section{Introduction}
\label{sec:intro}
\vspace{-0.3em}

Sound localization, the ability to determine a sound’s three- dimensional position, is vital in applications such as robotics, virtual reality, and human-computer interactions. It is especially crucial for CI users. CI users often struggle with accurate sound direction estimation compared to those with normal hearing \cite{ludwig2021sound}, which has significant implications for safety and their ability to navigate complex listening situations.

Traditional sound localization methods face two key challenges. Firstly, small microphone arrays face a significant performance decline, while large arrays are impractical for CI users. Secondly, prior studies on binaural sound localization have primarily concentrated on the frontal sphere, leaving the broader spectrum of sound localization largely unexplored. This research
explores the potential of DL to address these challenges
by learning complex features from binaural microphone signals while
achieving high accuracy.

HRTF describes how the human head and ears alter sound signals before they reach the eardrums.


Our research attempt to integrate DL techniques into sound localization utilizing a two-microphone setup and using the HRTF. This work proposes a model capable of accurately learning sound source localization from binaural recordings, especially for limited microphones or wearable devices such as hearing aids and CIs.

\section{Related Work}
\label{sec:related}

\subsection{Traditional sound localization and Duplex Theory}

Traditional research in sound localization primarily focused on understanding the role of interaural time differences (ITDs), interaural level differences (ILDs), and interaural phase differences (IPDs). Pioneering works by Rayleigh \cite{rayleigh1907dynamical} and Wallach \cite{wallach1940role} established the foundational importance of these parameters in the context of sound localization.
The Duplex Theory posits that there is a fundamental trade-off between IPD and ILD in sound source localization across different frequencies. At lower frequencies (below 1,360 Hz), minimal spectral differences exist due to wave diffraction caused by sound wave wavelengths exceeding 0.25 meters (for a head size of 0.25 meters at room temperature). Conversely, at frequencies above this threshold, more significant level differences are expected.
However, the upper frequency limit for detecting phase differences with two ears separated by 0.25 meters, is around 680 Hz. This limitation arises from the ambiguity in discerning which ear’s wave arrived first beyond a 180-degree phase difference. Consequently, IPD is more critical for lower frequencies, while ILD is more valuable for higher frequencies.

Middlebrooks’ work in 1991 \cite{middlebrooks1991sound} expanded our knowledge by highlighting spectral cues and the role of the pinna in sound elevation localization, bridging anatomical and acoustic aspects of the outer ear.

\vspace{-0.5em}

\subsection{Machine Learning Methods for Sound Localization}

In recent years, the field of sound localization has seen significant developments with a focus on the role of HRTF. Wightman and Kistler \cite{wightman1992dominant} used Principal Component Analysis (PCA) to customize HRTF for individual listeners, based on their specific anatomical characteristics. Talagala and Thushara \cite{talagala2012hrtf} introduced a novel approach by integrating various localization cues, including both time and phase differences, with spectral interaural differences.

Machine Learning (ML) has played a crucial role in enhancing sound localization accuracy. Early algorithms such as Gaussian Mixture Models (GMM) and Expectation–Maximization (EM) were employed \cite{woodruff2012binaural}\cite{schwartz2013speaker}.

Using DL for sound localization was done by Tsuzuki et al. \cite{tsuzuki2013approach} who used Multilayer-Perceptrons to estimate sound source localization from time delay and amplitude difference. Takeda et al. \cite{takeda2016sound} used Deep neural networks (DNNs) using phase information. Hirvonen et al. \cite{hirvonen2015classification} were the first to use feature extraction from spectrograms.
Recent DL advancements were also introduced by Chakrabarty \cite{chakrabarty2019multi}, Yang \cite{yang2021learning} and HU \cite{hu2023robust}.

The field of sound localization has been enriched by notable challenges, among which is the LOCATA Challenge, organized by the IEEE \cite{evers2020locata}.
This competition tackles sound localization and tracking using various microphone array setups.
It is noteworthy that the LOCATA challenge focuses on a single microphone array configuration based on binaural recordings and, interestingly, only one team chose to publish their work for this particular challenge \cite{augcaer2018binaural}.

Audio-Visual Correspondence, as demonstrated in “Look, Listen, and Learn” \cite{arandjelovic2017look} combines audio and visual data to acquire semantic knowledge. This approach processes audio and visual modalities separately with convolutional techniques and then combines them using dense layers. This approach had a distinct effect on the audio processing research.

The Hybrid Spectrogram and Waveform Source Separation method, introduced by Facebook AI Research \cite{defossez2021hybrid}, extends the U-Net model to handle both time and frequency domains in separating sources from hybrid waveform and spectrogram data.
Despite the emphasis on improving large microphone arrays, efficient techniques for limited microphone setups are essential, as demonstrated by challenges such as LOCATA.

\subsection{Benchmark model comparison}

Vecchiotti et al. \cite{vecchiotti2019end} introduced an influential method using CNNs to analyze waveform data directly for sound localization, diverging from traditional feature engineering. We use this work as a benchmark model for our comparison. Their model excelled in classifying sound sources location within a frontal 180-degree range among 37 speakers. We retrained it using our dataset and evaluated it using the same metrics.

The benchmark's model architecture includes a convolutional layer with linear activation function for frequency analysis, applied separately on both recording. Subsequent layers process both recordings, followed by two fully-connected hidden layers. The output layer uses softmax activation for sound localization classification, with training guided by the root mean square error (RMSE) loss function.

\section{Methodology}
\label{sec:methodology}

\subsection{Approach to address the challenge}
We adopted a hybrid approach that combines time and frequency domain data. Waveform data inherently contain temporal, spectral, and phase details. However, incorporating spectrogram information, has been shown to improve performance. Recent advances in DL, particularly in audio processing, have emphasized the utility of end-to-end systems for diverse applications.

Hence, we embraced an end-to-end hybrid model that leverages the strengths of both waveform and spectrogram representations. We believe that our model maximizes the benefits of recent advancements.

\subsection{Data collection and modeling }

Data collection occurred at the IRCAM studio in Paris, using the KU 100 dummy head microphone system. A precise arrangement of 24 strategically positioned speakers spanned three dimensions, with meticulous alignment using laser pointers for precision.

\vspace{-0.5em}
\begin{figure}[ht!]
\begin{minipage}[b]{.48\linewidth}
  \centering
\centerline{\includegraphics[width=4.0cm]{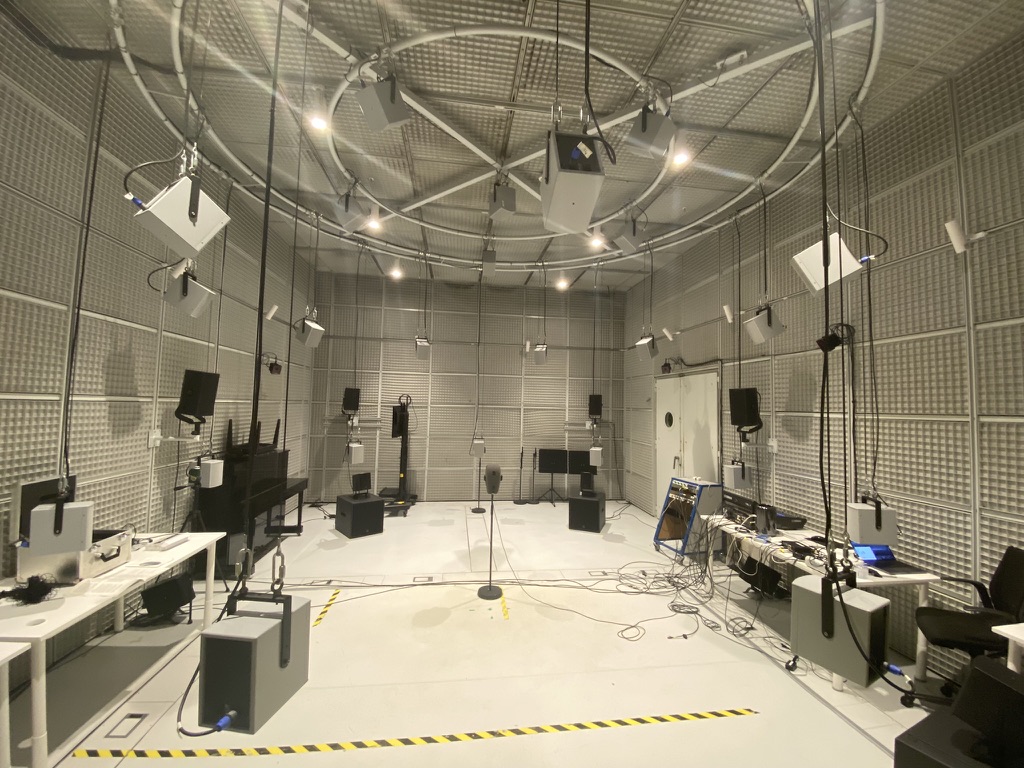}}
\end{minipage}
\hfill
\begin{minipage}[b]{0.48\linewidth}
  \centering
\centerline{\includegraphics[width=2.1cm]{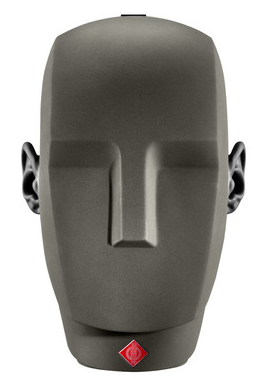}}
\end{minipage}
\vspace{-0.5em}

\caption{ Recording studio and KU100 Dummy head}
\end{figure}
\vspace{-0.5em}

Recordings involved generating “sweeps” across the frequency spectrum from 0 to 20,000 Hz, emitted from various speakers. These recordings were used to generate Head-Related Impulse Responses (HRIRs) for all directions and speakers. HRIRs are representations of how a sound impulse transforms in the time domain as it traverses from a defined spatial direction to reach a listener's ears.

Once HRIRs were established for each ear and sound source location, the recordings were convolved with these HRIRs. This convolution process allowed for the simulation of playback as if the recordings were emanating from specific spatial positions.

From the MUSDB18 dataset, 10 songs were randomly selected and convolutions were applied across 24 directions for both ears to create two datasets:

\begin{figure*}[t]
\centering
\includegraphics[width=0.8\textwidth]{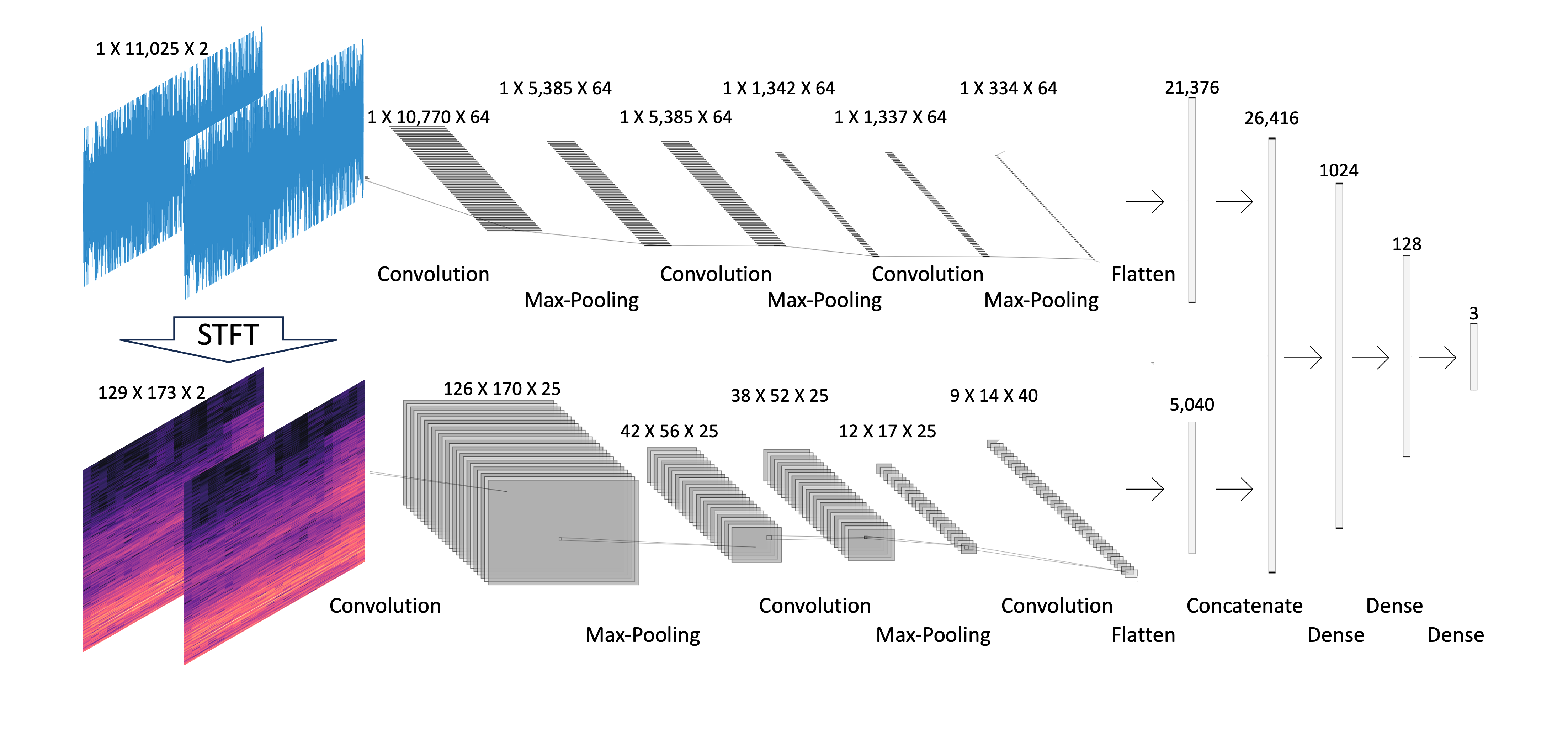}
\vspace{-3em}
\caption{\label{fig:model_arch}Hybrid time and frequency domain model architecture}
\end{figure*}

\begin{enumerate}
\item Raw Waveforms: This dataset consists of paired recordings from the left and right ears, concatenated along the channel dimension.
\item Spectrogram Dataset: We applied Short-Time Fourier Transform (STFT) on the waveform, using a 256-sample STFT window with a hop length of 64.
\end{enumerate}

To prevent data contamination, we applied these processes exclusively to five songs for the test dataset.

For GPU compatibility, we randomly selected 30\% of
the training data and 15\% of the test data while keeping the dataset size at 24 GB for training and evaluation.

\subsection{Architecture}

Our sound localization model (Figure 2) comprised of three key components: spectrogram unit, waveform unit and hybrid element that process the concatenation of both previous components through Relu activated DNN. The architecture is summarized in the following table.

\begin{table}[ht]
\centering
\small
\begin{tabular}{|c|c|c|}
\hline
Spectrogram & Waveform & Hybrid \\
\hline\hline
conv 4x4, 25 chan, Relu & conv 63, 75 chan, Relu &1024 units\\
\hline
max-pool 3x3, Relu & conv 59, 91 chan, tanh &128 units\\
\hline
conv 5x5, 25 chan, Relu & conv 58, 96 chan, tanh &3 units\\
\hline
2 max-pool 3x3, Relu & max-pool 10 & \\
\hline
conv 4x4, 40 chan, Relu & flatten & \\
\hline
\hline
\end{tabular}
\caption{Model's architecture}
\end{table}
\vspace{-1.5em}






\subsection{Loss Function and Experimental Configuration}
Previous studies \cite{tang2019regression} have shown that using Cartesian coordinates produces better results than polar coordinates, therefore, we selected this representation.

We initially used the MSE loss function but observed suboptimal convergence during the learning phase. To achieve a stable convergence and prioritize directional accuracy over distance accuracy, we used a sum of the Euclidean distance and the angular error.

\[Loss = \frac{1}{n}\left(\sum_{i=1}^{n} ||\textbf{x}_i-\textbf{x}'_i ||_2 + \frac{180}{\pi} \arccos(\sum_{i=1}^{n} \hat{\textbf{x}_i} \cdot \hat{\textbf{x}_{i}'})\right)\]




Where n is the number of instances, $\textbf{x}$ is the sound source location, $\textbf{x}'$ is the model prediction and $\hat{\textbf{x}}$ is the normalized vector location.

The model was trained using the Adam optimizer with a learning rate of $1 \times 10^{-3}$ and a batch size of 128 samples. The training process spanned 100 epochs, although models typically reached saturation between 40 to 70 epochs. A dynamic learning rate schedule was employed to enhance training.

\section{Results}

This section presents the outcomes achieved by our model, accompanied by a comprehensive analysis. We also present the performance of the benchmark model when applied to the same dataset and evaluation framework.

Figure 3 illustrates the average angular error associated with each speaker. The initial nine speakers are positioned on the lowest level, speakers 10 through 18 occupy the second level, speakers 19 to 23 are situated on the third level, and the 24th speaker is located precisely at the top of the spherical arrangement. Within each level, speakers are sequentially numbered starting from the front and proceeding in a clockwise direction.

Our model achieving an average angular error of $0.24^\circ$ and average Euclidean distance of 0.01 meters. Notably, there is no discernible consistent trend in the quality of results concerning either direction or frequency range. It is worth highlighting that there are two instances of directional outliers observed on the mid-level front side.

Figure 4 presents a similar metric, this time categorizing it according to frequency spans. The assignment of each data sample to a particular frequency range was determined by isolating the frequency component with the highest magnitude in that sample.  It is evident that there is no visible pattern across different frequency ranges. Furthermore, there is a notable outlier around the 3,000 Hz range, displaying a substantially larger angular error compared to other frequencies.

\begin{figure}[htb]
\centering
\includegraphics[width=0.7\linewidth]{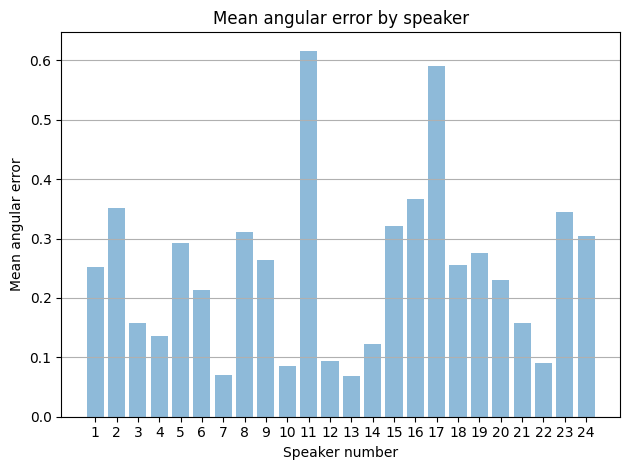}
\vspace{-1.5em}
\caption{Mean angular error in degrees for each speaker. The model' average angular error is $0.24^\circ$ while the benchmark's average angular error is $19.07^\circ$ }
\end{figure}

\vspace{-0.8em}
\begin{figure}[ht!]
\begin{minipage}[b]{\linewidth}
\centering
\centerline{\includegraphics[width=0.7\linewidth]{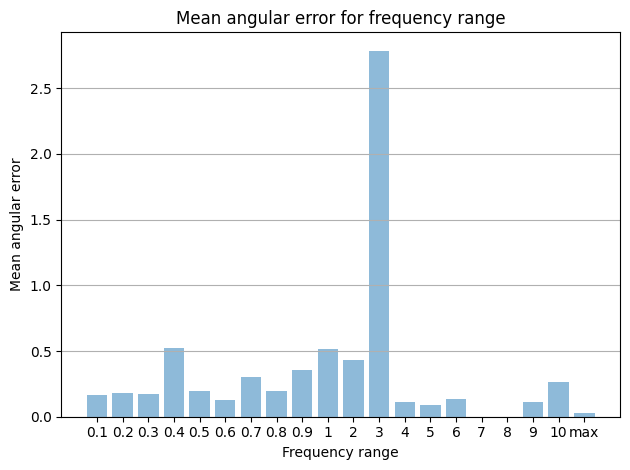}}
\end{minipage}
\vspace{-2.5em}
\caption{Mean angular error in degrees for each frequency range in kHz. We can see the high outlier at 3kHz of $2.7 ^\circ$ and the other frequencies below or around $0.5 ^\circ$}
\end{figure}

Figure 5 displays angular errors by speaker location. The second graph features 10 random samples, with blue vertical line denoting the actual localization and orange horizontal line representing the model's predictions. Notably, the model's predictions closely match the ground truth for all samples.

\vspace{-1.2em}
\begin{figure}[ht!]
\begin{minipage}[b]{.48\linewidth}
\centering
\centerline{\includegraphics[width=5cm]{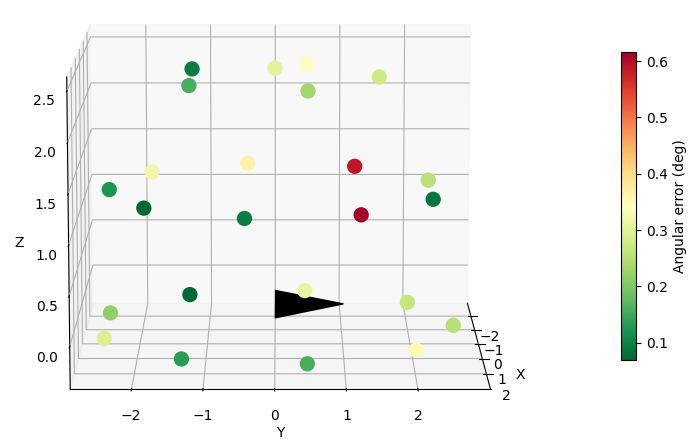}}
\end{minipage}
\hfill
\begin{minipage}[b]{0.48\linewidth}
\centering
\centerline{\includegraphics[width=4cm]{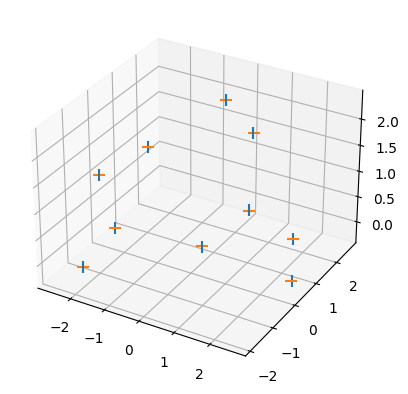}}
\end{minipage}
\vspace{-1em}
\caption{Left graph shows the mean angular error for each speaker. Right graph present example of source location (blue line) and the model's prediction (orange line).}
\label{fig:res}
\end{figure}
\vspace{-2em}

\subsection{Benchmark model results comparison}


The benchmark model results show an average angular error of $19.07^\circ$ and an average Euclidean distance of $1.08$ meters. Table 2 shows the comparison between our model and the benchmark model.

\vspace{-0.5em}
\begin{table}[ht!]
\centering
\small
\begin{tabular}{||c c c||} 
 \hline
  & Benchmark model & Hybrid model\\ [0.5ex] 
 \hline\hline
 Angular error & $19.07^\circ$ & $0.24^\circ$ \\ 
 \hline
 Euclidean distance & 1.08 m & 0.01 m \\ [0.5ex] 
 \hline
\end{tabular}
\vspace{-0.5em}
\caption{Hybrid Model and Benchmark model results}
\vspace{-0.5em}
\end{table}

Besides performance differences, the benchmark model shows distinctions in predicting frontal, rear, and above locations. Ablation studies using only waveform data on our model produced similar results, suggesting that incorporating frequency domain data enhances front-rear differentiation due to spectral cues from the pinna.

Figure 6 shows how the Benchmark model's performance declines as the horizontal angle increases, further supporting the hybrid time and frequency domain model's effectiveness.

\vspace{-0.6em}
\begin{figure}[ht!]
\centering
\includegraphics[width=0.9\linewidth]{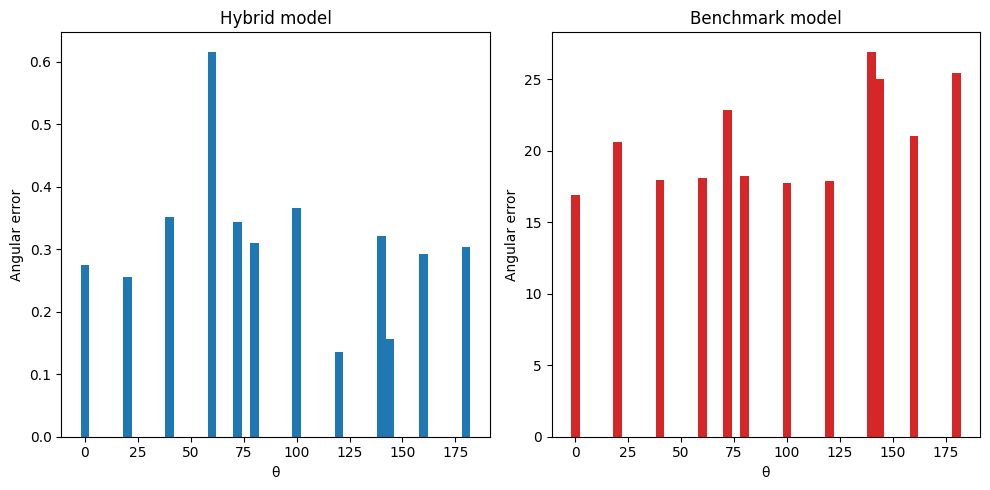}
\vspace{-1.5em}
\caption{Horizontal angular error by $\theta$ for hybrid and benchmark models. It is important to note both the magnitude of the angular error and how it trends concerning $\theta$}
\end{figure}
\vspace{-1.5em}

\section{Discussion and Conclusions}

Our study advances sound localization by extending its capabilities from a 180-degree range to a full-sphere context.
We introduce a hybrid model that outperforms existing methods.
Looking ahead, three promising research directions emerge:
\begin{enumerate}[noitemsep]

\item  Head-Agnostic, Configuration-Flexible Model: We aim to develop a model adaptable to diverse users and settings, eliminating the need for user-specific training.

\item Unified Model for Localization and Noise Cancelling: Integrating sound localization with noise cancellation or source separation could enhance both accuracy and noise handling.

\item Optimized Pinna Design: We believe it is possible to achieve reliable sound localization using a single microphone. This innovation could potentially reduce costs and enhance user comfort.
\end{enumerate}

We hope that these directives will enhance the quality of life for CI users and contribute to the advancement of various contemporary technologies.

\vspace{-1em}
\section*{Funding}
\vspace{-1em}

This research was funded by the European Research Council under the Horizon 2020 programme (Grant 883313 ERC REACH) and by the french national research agency ANR-19-CE23-0023 HAIKUS.

\newpage

\bibliographystyle{IEEEbib}
\bibliography{Paper_for_resubmission}

\begin{thebibliography}{10}

\bibitem{ludwig2021sound}
Alexandra~Annemarie Ludwig, Sylvia Meuret, Rolf-Dieter Battmer, Marc
  Sch{\"o}nwiesner, Michael Fuchs, and Arne Ernst,
\newblock ``Sound localization in single-sided deaf participants provided with
  a cochlear implant,''
\newblock {\em Frontiers in psychology}, vol. 12, pp. 753339, 2021.

\bibitem{rayleigh1907dynamical}
Lord Rayleigh,
\newblock ``On the dynamical theory of gratings,''
\newblock {\em Proceedings of the Royal Society of London. Series A, Containing
  Papers of a Mathematical and Physical Character}, vol. 79, no. 532, pp.
  399--416, 1907.

\bibitem{wallach1940role}
Hans Wallach,
\newblock ``The role of head movements and vestibular and visual cues in sound
  localization.,''
\newblock {\em Journal of Experimental Psychology}, vol. 27, no. 4, pp. 339,
  1940.

\bibitem{middlebrooks1991sound}
John~C Middlebrooks and David~M Green,
\newblock ``Sound localization by human listeners,''
\newblock {\em Annual review of psychology}, vol. 42, no. 1, pp. 135--159,
  1991.

\bibitem{wightman1992dominant}
Frederic~L Wightman and Doris~J Kistler,
\newblock ``The dominant role of low-frequency interaural time differences in
  sound localization,''
\newblock {\em The Journal of the Acoustical Society of America}, vol. 91, no.
  3, pp. 1648--1661, 1992.

\bibitem{talagala2012hrtf}
Dumidu~S Talagala and Thushara~D Abhayapala,
\newblock ``Hrtf aided broadband doa estimation using two microphones,''
\newblock in {\em 2012 International Symposium on Communications and
  Information Technologies (ISCIT)}. IEEE, 2012, pp. 1133--1138.

\bibitem{woodruff2012binaural}
John Woodruff and DeLiang Wang,
\newblock ``Binaural localization of multiple sources in reverberant and noisy
  environments,''
\newblock {\em IEEE Transactions on Audio, Speech, and Language Processing},
  vol. 20, no. 5, pp. 1503--1512, 2012.

\bibitem{schwartz2013speaker}
Ofer Schwartz and Sharon Gannot,
\newblock ``Speaker tracking using recursive em algorithms,''
\newblock {\em IEEE/ACM Transactions on Audio, Speech, and Language
  Processing}, vol. 22, no. 2, pp. 392--402, 2013.

\bibitem{tsuzuki2013approach}
Hirofumi Tsuzuki, Mauricio Kugler, Susumu Kuroyanagi, and Akira Iwata,
\newblock ``An approach for sound source localization by complex-valued neural
  network,''
\newblock {\em IEICE TRANSACTIONS on Information and Systems}, vol. 96, no. 10,
  pp. 2257--2265, 2013.

\bibitem{takeda2016sound}
Ryu Takeda and Kazunori Komatani,
\newblock ``Sound source localization based on deep neural networks with
  directional activate function exploiting phase information,''
\newblock in {\em 2016 IEEE international conference on acoustics, speech and
  signal processing (ICASSP)}. IEEE, 2016, pp. 405--409.

\bibitem{hirvonen2015classification}
Toni Hirvonen,
\newblock ``Classification of spatial audio location and content using
  convolutional neural networks,''
\newblock in {\em Audio Engineering Society Convention 138}. Audio Engineering
  Society, 2015.

\bibitem{chakrabarty2019multi}
Soumitro Chakrabarty and Emanu{\"e}l~AP Habets,
\newblock ``Multi-speaker doa estimation using deep convolutional networks
  trained with noise signals,''
\newblock {\em IEEE Journal of Selected Topics in Signal Processing}, vol. 13,
  no. 1, pp. 8--21, 2019.

\bibitem{yang2021learning}
Bing Yang, Hong Liu, and Xiaofei Li,
\newblock ``Learning deep direct-path relative transfer function for binaural
  sound source localization,''
\newblock {\em IEEE/ACM Transactions on Audio, Speech, and Language
  Processing}, vol. 29, pp. 3491--3503, 2021.

\bibitem{hu2023robust}
Qi~Hu, Ning Ma, and Guy~J Brown,
\newblock ``Robust binaural sound localisation with temporal attention,''
\newblock in {\em ICASSP 2023-2023 IEEE International Conference on Acoustics,
  Speech and Signal Processing (ICASSP)}. IEEE, 2023, pp. 1--5.

\bibitem{evers2020locata}
Christine Evers, Heinrich~W L{\"o}llmann, Heinrich Mellmann, Alexander Schmidt,
  Hendrik Barfuss, Patrick~A Naylor, and Walter Kellermann,
\newblock ``The locata challenge: Acoustic source localization and tracking,''
\newblock {\em IEEE/ACM Transactions on Audio, Speech, and Language
  Processing}, vol. 28, pp. 1620--1643, 2020.

\bibitem{augcaer2018binaural}
Semih A{\u{g}}caer and Rainer Martin,
\newblock ``Binaural source localization based on modulation-domain features
  and decision pooling,''
\newblock {\em arXiv preprint arXiv:1812.02399}, 2018.

\bibitem{arandjelovic2017look}
Relja Arandjelovic and Andrew Zisserman,
\newblock ``Look, listen and learn,''
\newblock in {\em Proceedings of the IEEE international conference on computer
  vision}, 2017, pp. 609--617.

\bibitem{defossez2021hybrid}
Alexandre D{\'e}fossez,
\newblock ``Hybrid spectrogram and waveform source separation,''
\newblock {\em arXiv preprint arXiv:2111.03600}, 2021.

\bibitem{vecchiotti2019end}
Paolo Vecchiotti, Ning Ma, Stefano Squartini, and Guy~J Brown,
\newblock ``End-to-end binaural sound localisation from the raw waveform,''
\newblock in {\em ICASSP 2019-2019 IEEE International Conference on Acoustics,
  Speech and Signal Processing (ICASSP)}. IEEE, 2019, pp. 451--455.

\bibitem{tang2019regression}
Zhenyu Tang, John~D Kanu, Kevin Hogan, and Dinesh Manocha,
\newblock ``Regression and classification for direction-of-arrival estimation
  with convolutional recurrent neural networks,''
\newblock {\em arXiv preprint arXiv:1904.08452}, 2019.

\end{thebibliography}

\end{document}